\documentclass[a4paper]{article}

\usepackage{INTERSPEECH2021}
\usepackage{booktabs}
\usepackage{caption}
\usepackage{subcaption}
\usepackage{multirow}
\usepackage{url}
\newcommand{\accessedDate}{Oct.~2020}
\newcommand{\urlfootnote}[1]{\footnote{\url{#1} Accessed \accessedDate}}

\title{Subjective Evaluation of Noise Suppression Algorithms in Crowdsourcing}
\name{Babak Naderi$^1$, Ross Cutler$^2$}
\address{
  $^1$Quality and Usability Lab, Technische Universit\"at Berlin\\
  $^2$Microsoft Corp.}
\email{babak.naderi@tu-berlin.de, ross.cutler@microsoft.com}

\begin{document}

\maketitle

\begin{abstract}
The quality of the speech communication systems, which include noise suppression algorithms, are typically evaluated in laboratory experiments according to the ITU-T Rec. P.835, in which participants rate background noise, speech signal, and overall quality separately. This paper introduces an open-source toolkit for conducting subjective quality evaluation of noise suppressed speech in crowdsourcing. We followed the ITU-T Rec. P.835, and P.808 and highly automate the process to prevent moderator's error.
To assess the validity of our evaluation method, we compared the Mean Opinion Scores (MOS), calculate using ratings collected with our implementation, and the MOS values from a standard laboratory experiment conducted according to the ITU-T Rec  P.835. Results show a high validity in all three scales namely background noise, speech signal and overall quality (average PCC = 0.961). Results of a round-robin test (N=5) showed that our implementation is also a highly reproducible evaluation method (PCC=0.99). Finally, we used our implementation in the INTERSPEECH 2021 Deep Noise Suppression Challenge \cite{reddy2021interspeech} as the primary evaluation metric, which demonstrates it is practical to use at scale. The results are analyzed to determine why the overall performance was the best in terms of background noise and speech quality. 
\end{abstract}
\noindent\textbf{Index Terms}: speech quality, crowdsourcing, P.835, noise suppression, subjective quality assessment

\section{Introduction}

Traditionally, the assessment of the speech quality, transmitted through telecommunication system, is commonly carried out by human test participants who are either instructed to hold a conversation over a telecommunication system under study (conversation test) or listen to short speech clips (listening-opinion tests) and afterwards rate perceived quality on one or several rating scales.
Speech calls can be carried out with various devices in different environments, commonly with a non-optimal acoustic surrounding. Therefore, noise suppression algorithms are widely integrated into the communication chain to enhance the quality of the speech communication system. 
Those systems are typically evaluated in laboratory-based listening tests according to the ITU-T Rec. P.835 \cite{ITU-P835} in which separate rating scales are used to independently estimate the quality of the \textit{Background noise}, the \textit{Speech Signal} and the \textit{Overall quality} alone.
Separate scales are used as the higher noise suppression level often adversely affects the speech or the signal component. Consequently, in a regular listening-only test, with a single-rating scale according to the ITU-T Rec. P.800, participants can often become confused as to what they should consider in rating the overall "quality". Accordingly, each individual determines her overall quality rating by weighting the signal and background components. Such a process introduces additional error variance in the overall quality ratings and reduces their reliability \cite{ITU-P835}. 

In addition, laboratory-based speech quality experiments are more and more replaced by crowdsourcing-based online tests, which are carried out by paid participants. 
Crowdsourcing offers a faster, scalable, and cheaper approach than traditional laboratory tests \cite{hossfeld2014best}. Still, with its challenges: the test participants take part in the test in their working environment using their hardware without a test moderator's direct supervision. 
Previous works showed that background noise in participant's surroundings can mask the degradation under the test and lead to a significantly different rating \cite{jimenez2020effect,naderi2018speech}. Different listening devices can also strongly influence the perceived quality \cite{ribeiro2011crowdmos}.
The ITU-T Rec, P.808 \cite{ITU-P808} addresses those challenges and provides methods to collect reliable and valid data in the crowdsourcing practice. However, the recommendation only focuses on the Absolute Category Rating (ACR) test method, whereas assessing the noise suppressed speech is more endangered by the environmental noise and uncalibrated listening device. In this work, we followed the methods described in the ITU-T Rec. P.808 and implemented the P.835 test procedure adapted to the crowdsourcing approach.
This work's contribution is as follows: we provide an open-source toolkit to conduct subjective assessment noise suppressed speech in crowdsourcing accessible to entire research and industry with no need of building specific laboratory. We show that our approach is highly valid, reliable, reproducible, and scalable in different experiments. We also provide insides about the state-of-the-art noise suppression algorithms' performance and point out the potential future direction in this domain.

This paper organized as following: Section 2 describes the toolkit's implementation and different components; Section 3 reports the validity and Section 4 the reproducibility studies we conducted;  Section 5 reports the evaluation of different models from INTERSPEECH 2021 DNS Challenge \cite{reddy2021interspeech} and the relation between the three scales used in the P.835 test. Finally, Section 6 discusses the findings and proposes steps for future work. 
\section{Implementation}
We have extended the open-source P.808 Toolkit \cite{naderi2020open} with methods for evaluating speech communication systems that include noise suppression algorithm\urlfootnote{https://github.com/microsoft/P.808}. We followed the ITU-T Rec. P.835 \cite{ITU-P835} and adapted it for the crowdsourcing approach based on the ITU-T Rec. P.808 \cite{ITU-P808}.

The P.808 Toolkit contains scripts for creating the crowdsourcing test (generate the HTML file, trapping stimuli, input URLs, etc.) and also a script for processing the submitted answers (i.e. data screening and aggregating the reliable ratings). We extended all components to support the P.835 test method.
The test includes several sections. In the \textit{qualification} section relevant demographic questions are asked and the hearing ability of test participants are examined using a digit-triplet test ~\cite{smits2004development}. In \textit{Setup} section, usage of both ear-pods and suitability of the participant's environment are evaluated using the modified just-noticeable difference in quality test method \cite{naderi2020env}. In the \textit{training} section, the test participant is introduced to the rating procedure and familiarized with the rating scale by rating a predefined set of stimuli. The stimuli in the training set should cover the entire range of the scales. The last section is the \textit{ratings} section in which the participant listens to a set of stimuli and rate them on the given scales. As the crowdsourcing task should be short, therefore, it is recommended to use about ten stimuli in the rating section. The participant can perform one or more tasks. The qualification section only appears once and if the participant passes the test it will not be shown in the next tasks. The setup and training sections appear once a while (every 30 and 60 minutes, respectively) when the worker passed them successfully.  
We kept the structure of the P.808 Toolkit the same; further details about the P.808 Toolkit can be found in \cite{naderi2020open} and details on validation of the ITU-T Rec. P.808 in \cite{naderi2020dtowards}. 

\begin{figure}[tb]
\centering
        \includegraphics[width=0.9\columnwidth]{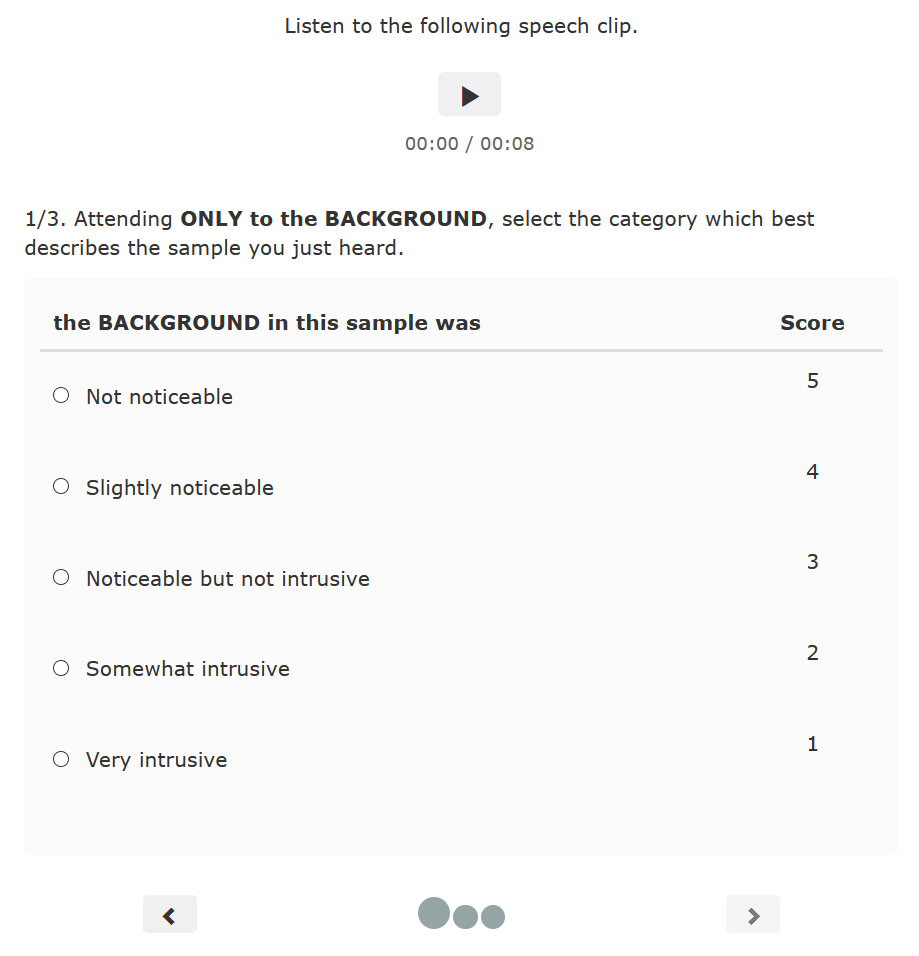}
        \caption{Screenshot of a trial in P.835 test as presented to the crowd workers.}
    \label{fig:sample}
\end{figure}

In the ITU-T Rec. P.835 subjective test, participants are ask to successively attend to and rate the stimulus on the speech signal (from \textit{1-Very distorted} to \textit{5-Not distorted}), the background noise (from \textit{1-Very intrusive} to \textit{5-Not noticeable}) and the overall quality (from \textit{1-Bad} to \textit{5-Excellent}) scales.
In our implementation, one clip is used for all three ratings as it was permitted by the recommendation \cite{ITU-P835}. Participants are forced to listen to the clip again before rating on each scale attending only to the scale's specific aspect\footnote{Although one time listening strongly reduces the working time on a task, our tests showed that it significantly influences the result.}. 
The presentation order of speech signal and background noise scales are randomized in each task to avoid any order effect. The overall quality scale is always the last rating in the sequence. Figure~\ref{fig:sample} illustrates the P835 trial as presented to the test participants.

In every crowdsourcing task, it is recommended to include trapping and gold questions \cite{naderi2018motivation, naderi2015effect}. The trapping stimuli is an obvious quality control mechanism \cite{naderi2015effect} which asks the participant to select a specific response to show their attention. For the P.835 extension, the trapping question asks the participant to select a specific score rather the category's label.

\subsection{Reference Conditions}
The reference conditions should be used in every subjective test, evaluating noisy speech, to independently vary the signal and background ratings through their entire range of scale values \cite{ITU-P835}. In the ITU-T Rec. P.835, Speech-to-Noise ratio (SNR) is used for varying the background noise (from 0 to 40 dB) and the Modulated  Noise  Reference Unit (MNRU) \cite{ITU-P810} for varying the signal rating (from 8 to 40 dBQ). Overall, 12 reference conditions are recommended. 
However, the previous body of works showed that MNRU processing is not appropriate as a reference system for the signal rating scale primary because the degradation in the speech signal by the noise canceller is very different from the degradation resulting from the MNRU processing \cite{AH-11-029}. Preliminary results from our crowdsourcing test and also expert review showed that software based MNRU degradation\footnote{We used software tools from ITU-T Rec. G.191 \cite{ITU-G191} to create the processing chain and apply degradations.} leads to a higher MOS rating than what is reported in the recommendation. 
Therefore, we applied the twelve reference conditions as propose in ETSI\,TS\,103\,281 \cite{ETSI_TS_103_281} (Table \ref{tab:rc}) in which the \textit{spectral subtraction based distortion} is used for degrading the speech signal. 
This signal distortion is based on the Wiener filter and leads to similar distortion to the one created by the noise cancellers. We used tools provided in \cite{S4_160397}, clean signals from \cite{ITU-P501} and noise signals from \cite{ETSI_TS_103_281} to create the reference conditions.

\begin{table}[tb]
    \caption{Reference conditions for fullband subjective evaluation of noise suppressors according to ETSI\,TS\,103\,281 \cite{ETSI_TS_103_281}.}
    \label{tab:rc} 
    \small
    \begin{center}
        \begin{tabular}{ l c c c}
        \toprule
        \textbf{Cond.} & \textbf{Speech}&		\textbf{SNR (A)} &	\textbf{Description}\\ 
        & \textbf{Distortion}& &\\
        \midrule
        i01 & - &  - & Best anchor for all\\
        i02 & -  &  0 dB & Lowest anchor for BAK\\
        i03 & -  &  12 dB & -\\
        i04 & - &  24 dB & - \\
        i05 & - & 36 dB & 2nd best anchor for BAK \\
        \midrule
        i06 & NS Level 1 &-& Lowest anchor for SIG\\
        i07 & NS Level 2 &-& -\\
        i08 & NS Level 3 &-& -\\
        i09 & NS Level 4 &-& 2nd best anchor for SIG\\
        \midrule
        i10 & NS Level 3 & 24 dB & 2nd best anchor for OVRL\\
        i11 & NS Level 2 & 12 dB &- \\
        i12 & NS Level 1 &   0 dB & Lowest anchor for OVRL\\
        \bottomrule
        \end{tabular}
    \end{center}
\end{table}
\section{Validation}

We conducted a crowdsourcing test using our P.835 extension of the P.808 Toolkit and compared its results with tests conducted in the laboratory according to the ITU-T Rec. P.835. 
In the test, we used four fullband speech files (2 male, 2 female) from the ITU-T  P.501  Annex  C \cite{ITU-P501} and applied the above-mentioned twelve reference conditions on them. 
On average we have collected 86.17 valid votes per test condition. 

\begin{figure*}[tb]
    \begin{subfigure}[c]{0.33\textwidth}
        \includegraphics[width=1.1\textwidth]{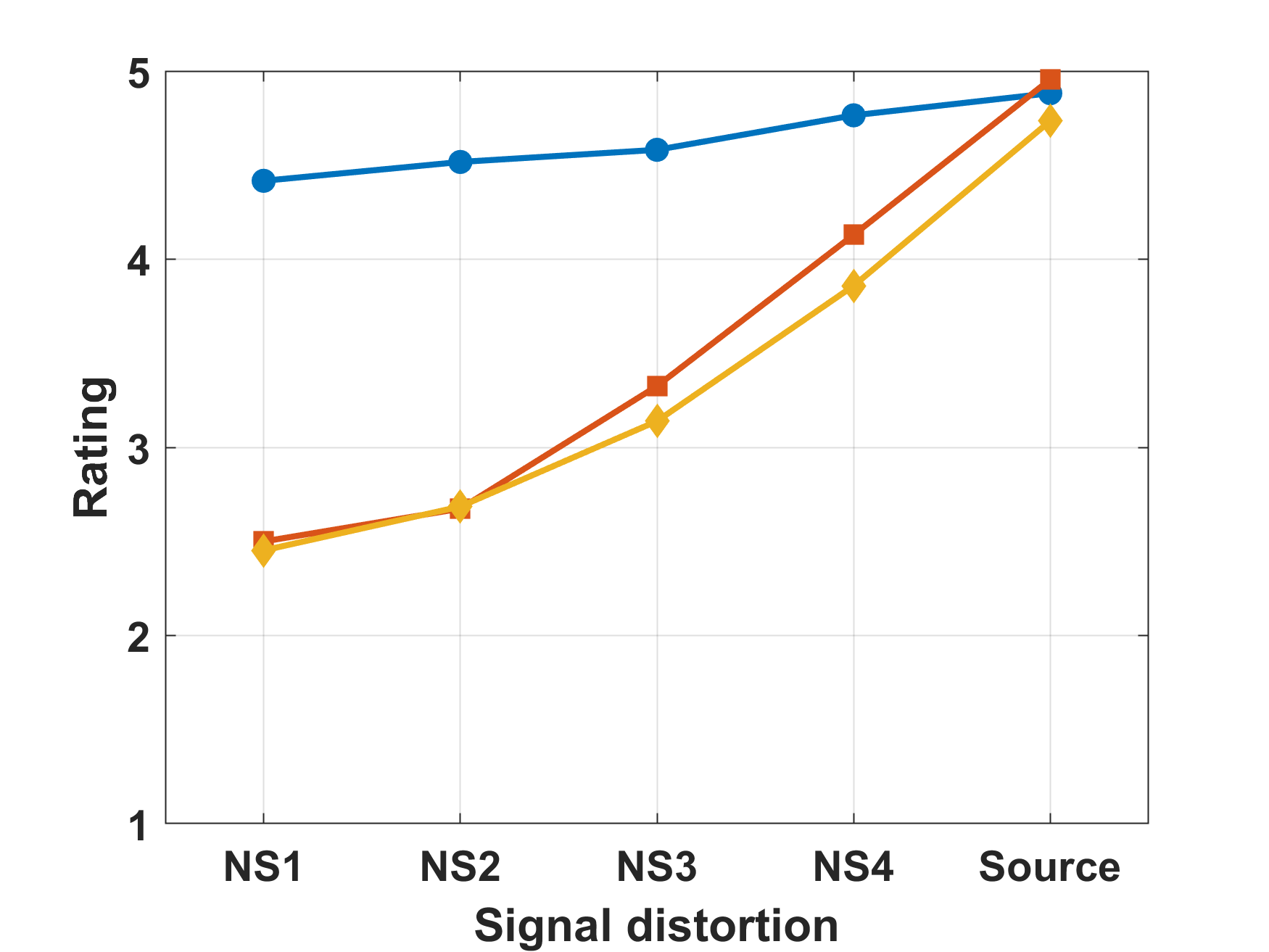}
        \caption{}
    \end{subfigure}    
    \begin{subfigure}[c]{0.33\textwidth}
        \includegraphics[width=1.1\textwidth]{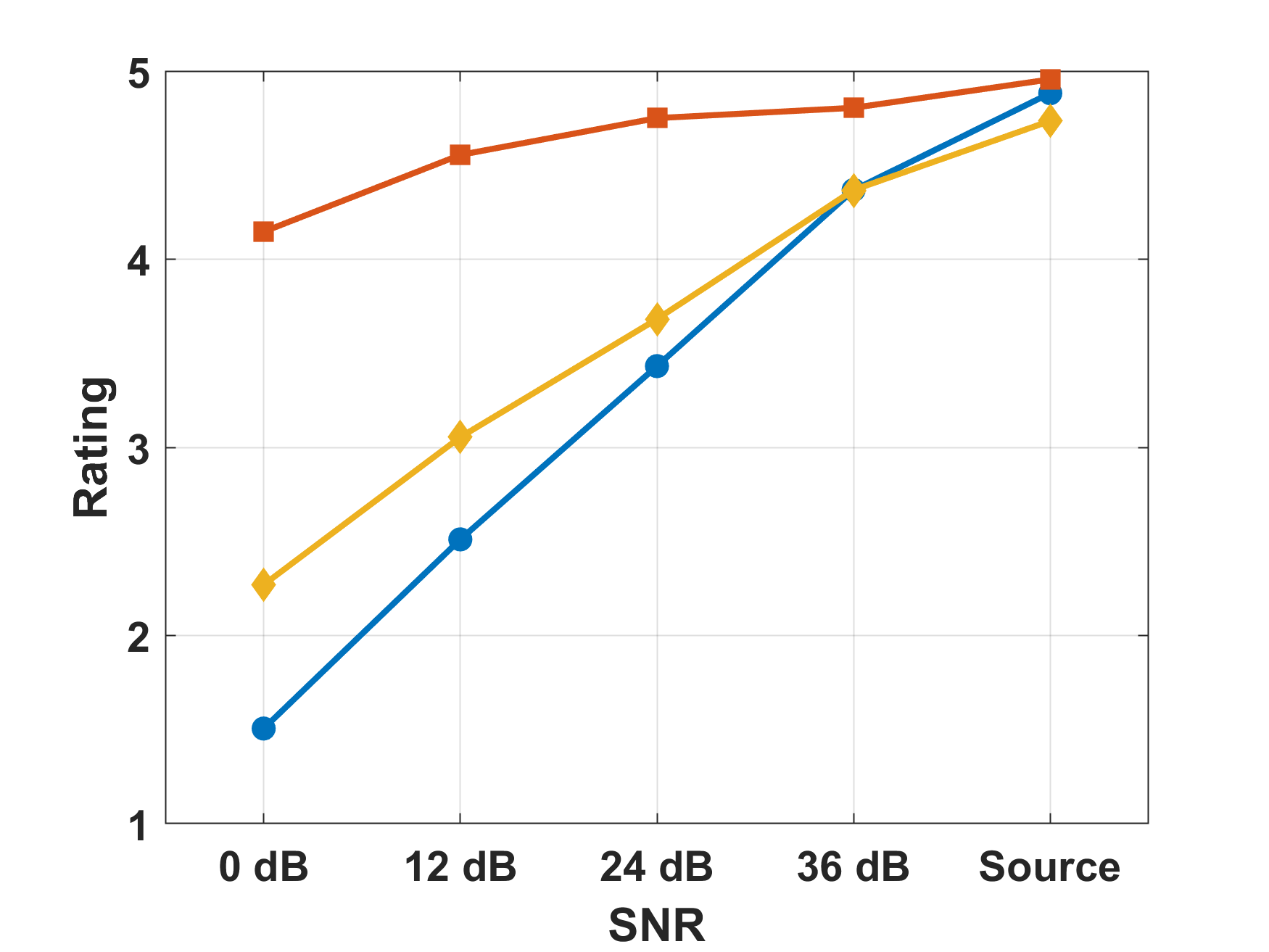}
        \caption{}
    \end{subfigure}   
    \begin{subfigure}[c]{0.33\textwidth}
        \includegraphics[width=1.1\textwidth]{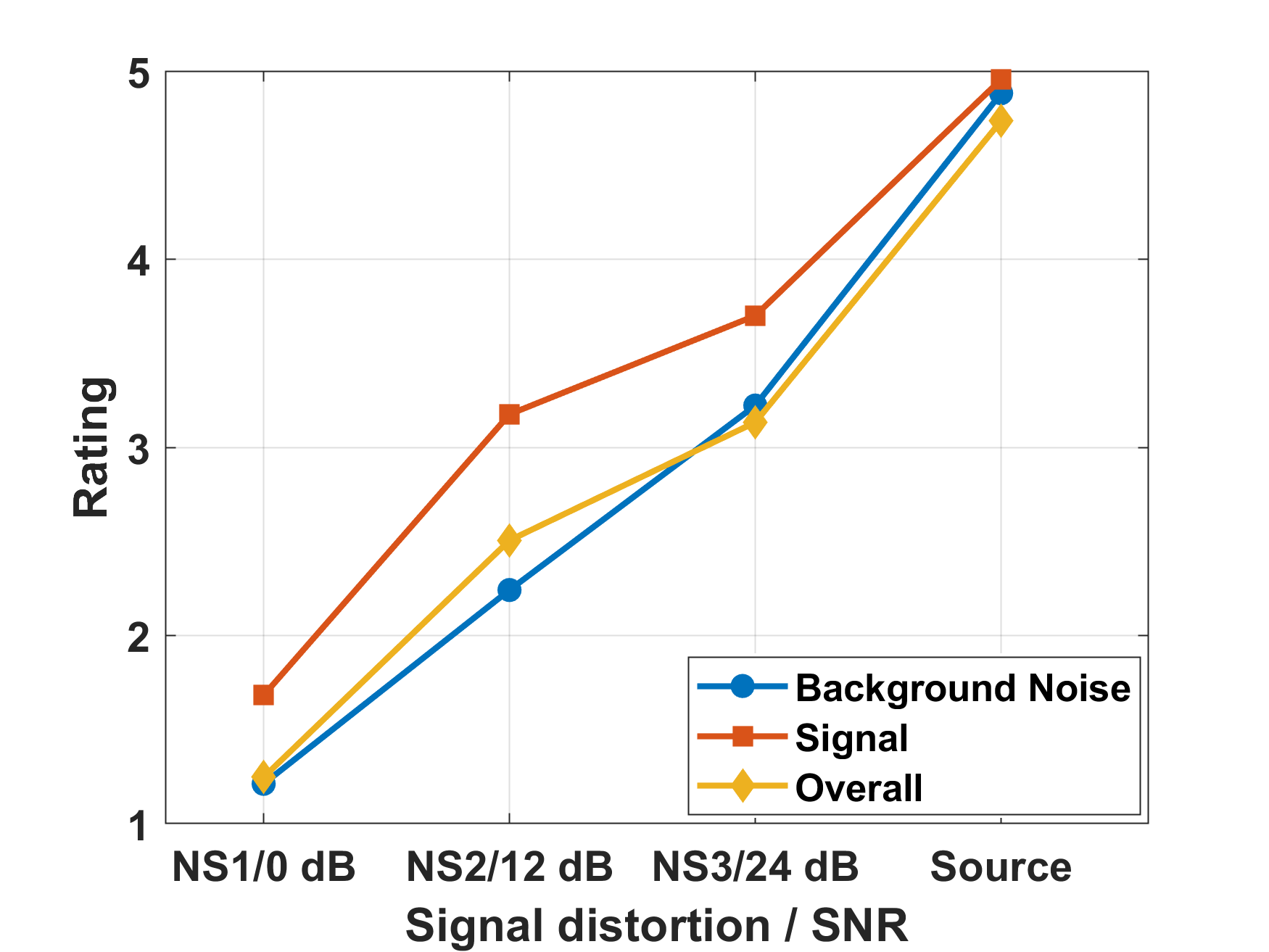}
        \caption{}
    \end{subfigure}    
    \caption{Auditory results of P.835 CS tests for the reference conditions. (a) No background noise, signal distortion varies, (b) background noise varies, signals not distorted, (c) signal distortion and background noise. Source referees to clean signal.}
    \label{fig:results:cs_ref_conditions}
\end{figure*}

\begin{table}[tbh]
    \caption{Comparison between Crowdsourcing (CS) and Laboratory (Lab) P.835 tests.}
    \label{tab:cs_vs_lab} 
    \begin{center}
    \resizebox{\columnwidth}{!}{%
        \begin{tabular}{ l c  c  c  c  }
        \toprule
        \textbf{Scale} &	\textbf{PCC}&	\textbf{RMSE}&	\textbf{Average }&\textbf{Average} \\ 
        &	\textbf{CS vs Lab}&	\textbf{CS vs Lab}&	\textbf{95\% CI CS}&\textbf{95\% CI Lab} \\ 
        \midrule
        Speech signal&	    0.925&	0.734&	0.17&	0.19 \\ 
        Background noise&	0.984&	0.507&	0.16&	0.11 \\ 
        Overall quality&	0.974&	0.33&	0.14&	0.18 \\ 
        \bottomrule
        \end{tabular}
        }
    \end{center}
\end{table}

Figure \ref{fig:results:cs_ref_conditions} illustrates the results of the crowdsourcing test. The overall quality ratings tend to be close to the minimum of signal and background noise rating.
Our results show high correlations to the openly available auditory results conducted in a laboratory from \cite{S4_150762} (c.f. Table \ref{tab:cs_vs_lab}).


\section{Reproducibility study}

In \cite{holub2017subjective}, the authors carried four subjective listening tests in three different laboratories to investigate inter- and intra-lab test result repeatability of P.835 methodology. The Pearson correlation between tests was high and above 0.97 in all cases. 

\begin{figure}[tb]
\centering
        \includegraphics[width=0.8\columnwidth]{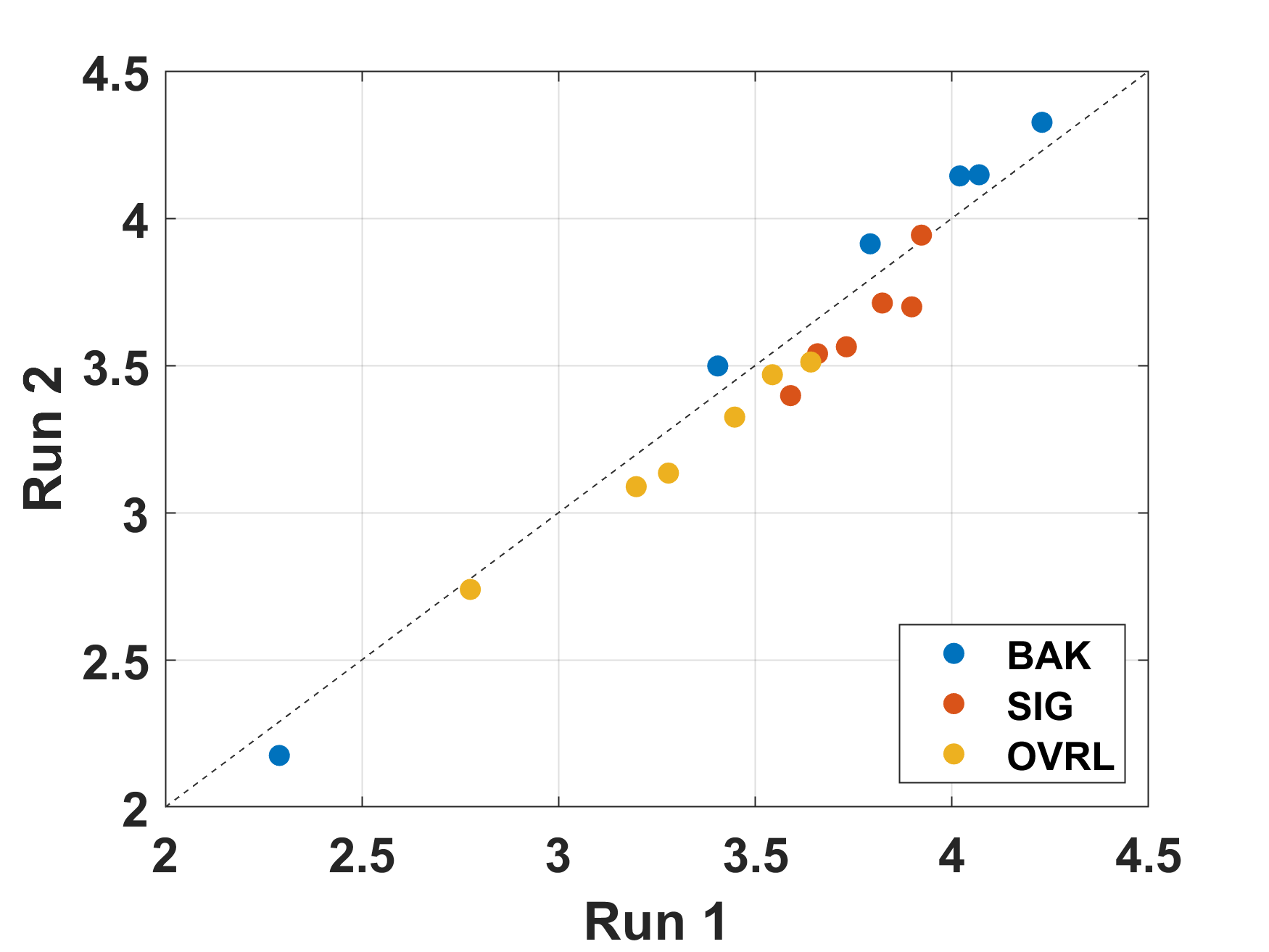}
        \caption{MOS values of the three scales from two runs in separate days.}
    \label{fig:results:repro}
\end{figure}

We used the blind test set from the Deep Noise Suppression (DNS) Challenge \cite{reddy2020interspeech} and applied 5 DNS models on that. We used the outcome and the unprocessed blind dataset for our reproducibility study.
The blind test set has 700 clips. A P.835 run was done N=5 times, on separate days and with mutually exclusive raters. The results are shown in Figure \ref{fig:results:repro} for N=2 runs, and show very good reproducibility (c.f. Table \ref{tab:reproducibility} for N=5 runs).

\begin{table}[tb]
\caption{P.835 reproducibility for N=5 runs. Rank transformation is recommended in case of small number of conditions for Spearman correlation\cite{naderi2020mos}.}
\label{tab:reproducibility}
\begin{center}
\resizebox{\columnwidth}{!}{%
\begin{tabular}{c c c c c c c c c c c}
\toprule
\multicolumn{3}{c}{\textbf{PCC}} & & \multicolumn{3}{c}{\textbf{SRCC}} & &
\multicolumn{3}{c}{\textbf{SRCC trans. rank}}\\
OVLR & BAK & SIG & & OVLR & BAK & SIG  & & OVLR & BAK & SIG \\
\midrule
0.99 & 0.99	& 0.99 & & 1.00 & 1.00 & 0.94 & &
0.99 & 0.94 & 0.97\\
\bottomrule
\end{tabular}
}
\end{center}

\end{table}

\section{INTERSPEECH 2021 Deep Noise Suppression Challenge}
Our P.835 tool was used in the the INTERSPEECH 2021 Deep Noise Suppression Challenge \cite{reddy2021interspeech} as the primary evaluation metric. The results of P.835 evaluation for the teams in Track 1 are reported in Table \ref{tab:P835}. Each team entry was evaluated with $N \sim 2600$ ratings and the resulting 95\% confidence interval was 0.04 for MOS values of \textit{BAK}, \textit{SIG}, and \textit{OVRL}.

Hu et al. \cite{hu2007subjective} estimated the relationship between the background noise (\textit{BAK}), the speech signal (\textit{SIG}), and the overall quality (\textit{OVRL}) as following using the NOIZEUS dataset.
\begin{equation}
\label{eq:0}
\scriptstyle  \widehat{OVRL}_{MOS}\,=\,-0.0783\,+\,0.571\,SIG_{MOS}\,+\,0.366\,BAK_{MOS}
\end{equation}

Using the challenge results and the reference conditions we determined the following prediction equation applying linear regression: 
\begin{equation}
\label{eq:1}
\scriptstyle \widehat{OVRL}_{MOS}\,=\,-0.844 \,+\, 0.644 \,SIG_{MOS}  \,+\, 0.452\, BAK_{MOS} 
\end{equation}

\noindent with adjusted $R^2=0.98$ and $\rho=0.98$, which is similar to the relation \cite{hu2007subjective} estimated with a different dataset. In particular $SIG_{MOS} /  BAK_{MOS} = 1.56$ for Eq. \ref{eq:0} and $1.42$ for Eq. \ref{eq:1}, showing that the signal quality has more weight for overall quality compared to the noise quality.

We applied both Eq.\ref{eq:0} and \ref{eq:1} to the data we collected in Section 3. The predicted values highly correlate with the collected data (PCC=0.97, RMSE=0.51 for Eq. \ref{eq:0}, and PCC=0.96, RMSE=0.27 for Eq. \ref{eq:1}).

\begin{figure}[tbh]
\centering
        \includegraphics[width=0.9\columnwidth]{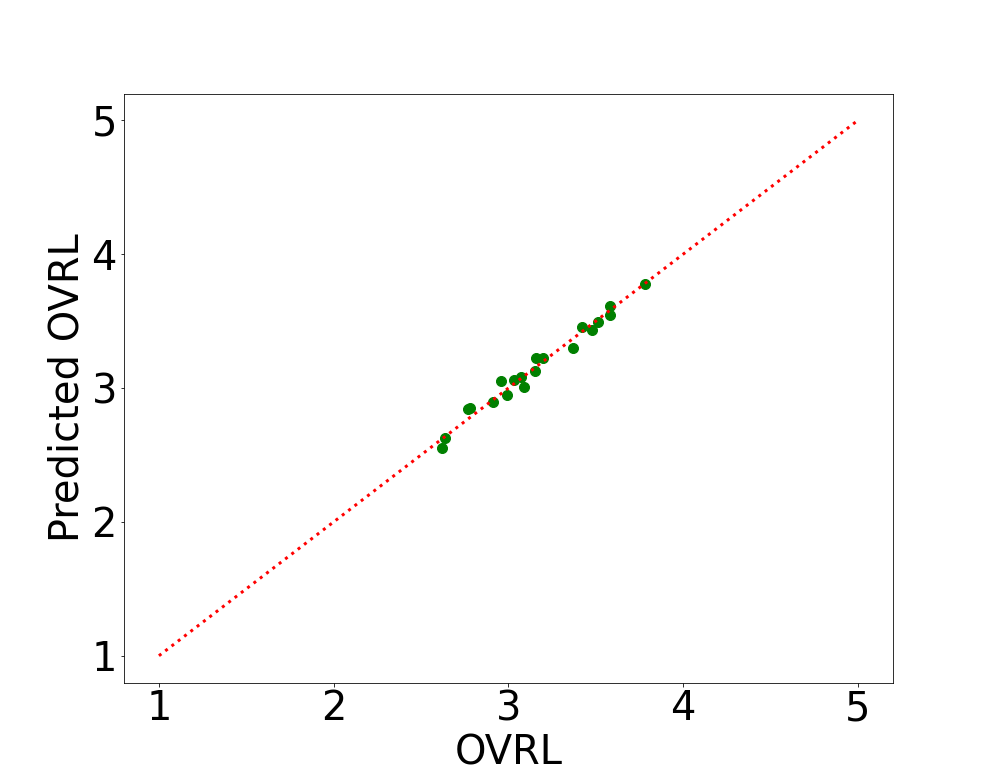}
        \caption{Linear regression of $OVRL_{MOS} \sim SIG_{MOS} + BAK_{MOS}$. Red line is ideal fit.}
    \label{fig:linear_regression}
\end{figure}

The challenge results show that the best overall performing model, \textit{Team 36}, achieves Difference Mean Opinion Score (DMOS) of 1.01 by very significant noise suppression (BAK\textsubscript{DMOS}=2.05) and with no extra speech degradation (SIG\textsubscript{DMOS}=0.01). There is an additional BAK\textsubscript{DMOS}=0.22 improvement that can be made in noise suppression (comparing to the reference condition i01, i.e., no degradation, with BAK\textsubscript{MOS}=4.88). Using the Eq. \ref{eq:1} for BAK\textsubscript{MOS}=5 and SIG\textsubscript{MOS}=3.89, the predicted OVLR\textsubscript{MOS} is 3.92, which is an estimate of the best DNS we can do without improving signal quality. To get significantly better performance the speech signal must be improved, such as through dereverberation or capture device distortion removal.

\begin{table}[tbh]
\caption{P.835 results from INTERSPEECH 2021 Deep Noise Suppression Challenge Track 1}
\label{tab:P835}
\begin{center}
\resizebox{\columnwidth}{!}{
    \begin{tabular}{l c c c c c c c}
    \toprule
     \multirow{2}{*}{\textbf{Team}} & \multicolumn{3}{c}{\textbf{MOS}} && \multicolumn{3}{c}{\textbf{DMOS}}\\
     & BAK & SIG& OVRL & &  BAK & SIG& OVRL \\
    \midrule
    36 & 4.66 & 3.90 & 3.78 && 2.05 & 0.01 & 1.01 \\
    33 & 4.48 & 3.77 & 3.58 && 1.87 & -0.12 & 0.81 \\
    13 & 4.35 & 3.76 & 3.58 && 1.74 & -0.13 & 0.80 \\
    34 & 4.29 & 3.72 & 3.51 && 1.68 & -0.17 & 0.74 \\
    19 & 4.13 & 3.74 & 3.48 && 1.52 & -0.15 & 0.71 \\
    18 & 4.52 & 3.50 & 3.42 && 1.91 & -0.39 & 0.64 \\
    16 & 3.76 & 3.79 & 3.37 && 1.16 & -0.10 & 0.60 \\
    8 & 4.20 & 3.37 & 3.20 && 1.59 & -0.52 & 0.42 \\
    22 & 4.34 & 3.27 & 3.16 && 1.73 & -0.62 & 0.39 \\
    20 & 3.89 & 3.44 & 3.15 && 1.28 & -0.45 & 0.38 \\
    31 & 3.73 & 3.36 & 3.09 && 1.12 & -0.53 & 0.32 \\
    baseline & 3.89 & 3.36 & 3.07 && 1.28 & -0.54 & 0.30 \\
    12 & 4.07 & 3.20 & 3.03 && 1.47 & -0.69 & 0.25 \\
    30 & 3.46 & 3.46 & 2.99 && 0.85 & -0.43 & 0.22 \\
    37 & 4.18 & 3.11 & 2.96 && 1.58 & -0.78 & 0.19 \\
    11 & 3.81 & 3.13 & 2.91 && 1.20 & -0.76 & 0.14 \\
    38 & 2.59 & 3.92 & 2.78 && -0.02 & 0.03 & 0.01 \\
    noisy & 2.61 & 3.89 & 2.77 && 0.00 & 0.00 & 0.00 \\
    28 & 3.60 & 2.86 & 2.64 && 1.00 & -1.03 & -0.13 \\
    4 & 2.84 & 3.28 & 2.62 && 0.23 & -0.61 & -0.15 \\
    \bottomrule
    \end{tabular}
}
\end{center}
\end{table}

\section{Discussion and Conclusion}

We have provided an open-source toolkit for evaluating noise suppression algorithms using subjective tests conducted in the crowdsourcing approach. We followed the ITU-T Rec. P.835 and P.808 and applied the recommended hearing, device usage, and environment suitability tests, as well as the gold standard and trapping questions in the crowdsourcing tasks to ensure the reliability of the estimations.
We provide our P.835 implementation as an extension for the P.808 Toolkit.  Following the same structure, the toolkit is highly automated to avoid operational errors. 
The toolkit makes conducting subjective assessment according to the ITU-T recommendations accessible for the entire research and industry community without any extra cost of building the specific laboratory.
We conducted a validity study in which we observed high correlations between MOS values of the three scales calculated on ratings collected by our toolkit and the MOS values from the standard ITU-T Rec. P.835 laboratory-based tests (average PCC = 0.961).

We examined the reproducibility of the subjective ratings collected by our implementation. We collected ratings for 4200 clips, including clips processed by five DNS models and the unprocessed ones, in five runs in separate days with mutually exclusive raters. Results show a very good reproducibility (average PCC = 0.98, SPCC = 0.98, SPCC after transformation = 0.97).  

We also evaluated the results from the INTERSPEECH 2021 Deep Noise Suppression Challenge \cite{reddy2020icassp} using our P.835 implementation. Results show that the best performing model increases the background noise quality by 2.05 MOS without adding extra distortion to the signal quality. Consequently, significant improvement in the performance of the state-of-the-art DNS models can only be achieved by speech signal enhancement. Such a large scale evaluation, in which we collected about 78,000 votes for each rating scale, also demonstrates the scalability of our approach even during the pandemic. 

We observed that the P.835 test duration is significantly longer than the P.808 ACR test, as participants need to listen to the speech clips three times. The cost of the P.835 increases by $2\times$ compared to P.808, which is less than $3\times$ due to the common qualification overhead in both P.808 and P.835. 
For future work, continuous environment monitoring might be considered, which only exposes the participants to a new environment suitability test when a substantial change in the environment is detected. That can strongly decrease the test duration by reducing the extra overheads per session. Meanwhile, the application of continuous rating scales rather than discrete ones should be evaluated considering whether they provide higher sensitivity measurements.

\bibliographystyle{IEEEbib}
\bibliography{strings,refs}

\begin{thebibliography}{10}

\bibitem{reddy2021interspeech}
Chandan~KA Reddy, Harishchandra Dubey, Kazuhito Koishida, Arun Nair, Vishak
  Gopal, Ross Cutler, Sebastian Braun, Hannes Gamper, Robert Aichner, and
  Sriram Srinivasan,
\newblock ``{INTERSPEECH} 2021 deep noise suppression challenge,''
\newblock in {\em INTERSPEECH}, 2021.

\bibitem{ITU-P835}
{ITU-T Recommendation P.835},
\newblock {\em {Subjective test methodology for evaluating speech communication
  systems that include noise suppression algorithm}},
\newblock International Telecommunication Union, Geneva, 2003.

\bibitem{hossfeld2014best}
Tobias Ho{\ss}feld, Matthias Hirth, Judith Redi, Filippo Mazza, Pavel
  Korshunov, Babak Naderi, Michael Seufert, Bruno Gardlo, Sebastian Egger, and
  Christian Keimel,
\newblock ``Best practices and recommendations for crowdsourced qoe-lessons
  learned from the qualinet task force" crowdsourcing",''
\newblock 2014.

\bibitem{jimenez2020effect}
Rafael~Zequeira Jim{\'e}nez, Babak Naderi, and Sebastian M{\"o}ller,
\newblock ``Effect of environmental noise in speech quality assessment studies
  using crowdsourcing,''
\newblock in {\em 2020 Twelfth International Conference on Quality of
  Multimedia Experience (QoMEX)}. IEEE, 2020, pp. 1--6.

\bibitem{naderi2018speech}
Babak Naderi, Sebastian M{\"o}ller, and Gabriel Mittag,
\newblock ``Speech quality assessment in crowdsourcing: Influence of
  environmental noise,''
\newblock in {\em 44. Deutsche Jahrestagung f{\"u}r Akustik (DAGA)}, pp.
  229--302. Deutsche Gesellschaft f{\"u}r Akustik DEGA eV, 2018.

\bibitem{ribeiro2011crowdmos}
Fl{\'a}vio Ribeiro, Dinei Flor{\^e}ncio, Cha Zhang, and Michael Seltzer,
\newblock ``Crowdmos: An approach for crowdsourcing mean opinion score
  studies,''
\newblock in {\em 2011 IEEE international conference on acoustics, speech and
  signal processing (ICASSP)}. IEEE, 2011, pp. 2416--2419.

\bibitem{ITU-P808}
{ITU-T Recommendation P.808},
\newblock {\em {Subjective evaluation of speech quality with a crowdsourcing
  approach}},
\newblock International Telecommunication Union, Geneva, 2018.

\bibitem{naderi2020open}
Babak Naderi and Ross Cutler,
\newblock ``An open source implementation of itu-t recommendation p.808 with
  validation,''
\newblock in {\em INTERSPEECH}. 2020, ISCA.

\bibitem{smits2004development}
Cas Smits, Theo~S Kapteyn, and Tammo Houtgast,
\newblock ``Development and validation of an automatic speech-in-noise
  screening test by telephone,''
\newblock {\em International journal of audiology}, vol. 43, no. 1, pp. 15--28,
  2004.

\bibitem{naderi2020env}
Babak Naderi and Sebastian M{\"o}ller,
\newblock ``{Application of Just-Noticeable Difference in Quality as
  Environment Suitability Test for Crowdsourcing Speech Quality Assessment
  Task},''
\newblock in {\em 2020 Twelfth International Conference on Quality of
  Multimedia Experience (QoMEX)}, 2020, pp. 1--6.

\bibitem{naderi2020dtowards}
Babak Naderi, Rafael~Zequeira Jiménez, Matthias Hirth, Sebastian M\"oller,
  Florian Metzger, and Tobias Ho\ss{}feld,
\newblock ``Towards speech quality assessment using a crowdsourcing approach:
  Evaluation of standardized methods,''
\newblock {\em Quality and User Experience}, vol. 5, no. 1, pp. 1--21, 2020.

\bibitem{naderi2018motivation}
Babak Naderi,
\newblock {\em Motivation of workers on microtask crowdsourcing platforms},
\newblock Springer, 2018.

\bibitem{naderi2015effect}
Babak Naderi, Tim Polzehl, Ina Wechsung, Friedemann K{\"o}ster, and Sebastian
  M{\"o}ller,
\newblock ``{Effect of Trapping Questions on the Reliability of Speech Quality
  Judgments in a Crowdsourcing Paradigm},''
\newblock in {\em Conference of the International Speech Communication
  Association}, 2015.

\bibitem{ITU-P810}
{ITU-T Recommendation P.810},
\newblock {\em {Modulated noise reference unit (MNRU)}},
\newblock International Telecommunication Union, Geneva, 1996.

\bibitem{AH-11-029}
{AH-11-029},
\newblock {\em {Better Reference System for the P.835 SIG Rating Scale}},
\newblock Rapporteur's meeting, International Telecommunication Union, Geneva,
  Switzerland, 20-21 June 2011.

\bibitem{ITU-G191}
{ITU-T Recommendation G.191},
\newblock {\em {Software tools for speech and audio coding standardization}},
\newblock International Telecommunication Union, Geneva, 2019.

\bibitem{ETSI_TS_103_281}
{ETSI TS 103 281 v1.3.1},
\newblock {\em {Speech and multimedia Transmission Quality (STQ); Speech
  quality in the presence of background noise: Objective test methods for
  super-wideband and fullband terminals }},
\newblock ETSI, France, 2019.

\bibitem{S4_160397}
{S4‑160397},
\newblock {\em {Revision of DESUDAPS-1: Common subjective testing framework for
  training and validation of SWB and FB P.835 test predictors v 1.2 }},
\newblock 3GPP, Memphis, TN, USA, 2016.

\bibitem{ITU-P501}
{ITU-T Recommendation P.501},
\newblock {\em {Test signals for use in telephony and other speech-based
  applications}},
\newblock International Telecommunication Union, Geneva, 2020.

\bibitem{S4_150762}
{S4‑150762},
\newblock {\em {Reference Impairments for Superwideband and Fullband P.835
  Tests – Processing and Auditory Results }},
\newblock 3GPP, Rennes, France, 2015.

\bibitem{holub2017subjective}
Jan Holub, Hakob Avetisyan, and Scott Isabelle,
\newblock ``Subjective speech quality measurement repeatability: comparison of
  laboratory test results,''
\newblock {\em International Journal of Speech Technology}, vol. 20, no. 1, pp.
  69--74, 2017.

\bibitem{reddy2020interspeech}
Chandan~KA Reddy, Vishak Gopal, Ross Cutler, Ebrahim Beyrami, Roger Cheng,
  Harishchandra Dubey, Sergiy Matusevych, Robert Aichner, Ashkan Aazami,
  Sebastian Braun, et~al.,
\newblock ``The interspeech 2020 deep noise suppression challenge: Datasets,
  subjective testing framework, and challenge results,''
\newblock {\em arXiv preprint arXiv:2005.13981}, 2020.

\bibitem{naderi2020mos}
Babak Naderi and Sebastian Möller,
\newblock ``Transformation of mean opinion scores to avoid misleading of ranked
  based statistical techniques,''
\newblock in {\em 2020 Twelfth International Conference on Quality of
  Multimedia Experience (QoMEX)}, 2020, pp. 1--4.

\bibitem{hu2007subjective}
Yi~Hu and Philipos~C Loizou,
\newblock ``Subjective comparison and evaluation of speech enhancement
  algorithms,''
\newblock {\em Speech communication}, vol. 49, no. 7-8, pp. 588--601, 2007.

\bibitem{reddy2020icassp}
Chandan~KA Reddy, Harishchandra Dubey, Vishak Gopal, Ross Cutler, Sebastian
  Braun, Hannes Gamper, Robert Aichner, and Sriram Srinivasan,
\newblock ``{ICASSP} 2021 deep noise suppression challenge,''
\newblock in {\em ICASSP}, 2021.

\end{thebibliography}

\end{document}